\RequirePackage{fix-cm}
\documentclass[smallextended]{svjour3}       
\smartqed  
\usepackage{graphicx}
 \journalname{Biophysical Economics and Sustainability}
\begin{document}

\title{Growth and inequalities in a physicist's view
}


\author{Angelo Tartaglia
}


\institute{INAF and Politecnico di Torino \at
              Corso Duca degli Abruzzi 24, 10129 Torino, Italy \\
              Tel.: +39-011-0907328\\
              Fax: +39-011-0907399\\
              \email{angelo.tartaglia@polito.it}           
}

\date{Received: date / Accepted: date}

\maketitle

\begin{abstract}
It is still common wisdom amongst economists, politicians and lay people that economic growth
is a necessity of our social systems, at least to avoid distributional conflicts. This paper challenges
such belief moving from a purely physical theoretical perspective. It formally considers the
constraints imposed by a finite environment on the prospect of continuous growth, including the
dynamics of costs. As costs grow faster than production it is easy to deduce a final unavoidable
global collapse. Then, analyzing and discussing the evolution of the unequal share of wealth under
the premises of growth and competition, it is shown that the increase of inequalities is a necessary
consequence of the premises.
\end{abstract}

\section{Introduction}
\label{intro}
A tenet of contemporary economy is the need for growth and an old
open question is about its global sustainability over an indefinite future.
On one side ordinary common sense should immediately suggest that infinite
growth in a finite environment is impossible, on the other one should
clarify what kind of "growth" we are speaking about. The issue has been
debated since long time ago, at least for some peculiar aspects as, mainly,
the world population. Thomas Malthus \cite{malthus} focused on the long term consequences of a
mismatch between the growth rates of the human kind and of the food
production. Apparently no worry was presented on the possibility that
primary resources could globally become exhausted, probably under the
feeling that their amount was \textit{practically} infinite; an attitude
understandable at the end of the $18^{th}$ century, but totally
unjustified in today's globalized economy. The problem of growth, still
referred to the world population, was faced again fifty years later, by
Pierre Verhulst \cite{verhulst}, presenting the only possible growth law in
limited availability of food conditions.

The very problem of global sustainability of growth, at least in the
attention of the general public, was discussed, more than one century after
the above quoted examples, in \textit{The limits to growth }\cite{limits}
authored by Donella Meadows, Dennis Meadows, J{\o}rgen Randers and William
Behrens III on behalf of a research group at MIT. The study had been
commissioned by the Club of Rome and was based on the then novel approach of
\textit{System Dynamics} applied to the world economy as a complex system of
inter-human relations embedded into a wider material complex system
corresponding to the environment where humans live in. As soon as the report appeared it was harshly criticized by many academic, economical and political milieus, without caring about the soundness of its hypotheses but rejecting its unwelcome conclusions. As for the subsequent debate up to this date, the impression one obtains is that the attention is mostly concentrated on empirical and behavioral features of subsystems (one country or region rather than another), leaving in an unattended background any idea of material constraints \cite{wei,helpman,brenner}. Remarkable attempts to include the limits into global sophisticated models also exist, but in general they are worked out starting from domains primarily pertaining to fields of knowledge external to economics. An example is a theory elaborated by Tim Garrett \cite{garrett} linking the CO$_2$ emissions problem to the joint dynamics of the global material and human systems. The paradigm of thermodynamics is applied to describe the material functioning of any human society, re-interpreting growth dynamics and collapse in terms of energy (and entropy) flow. The idea is interesting and has been taken up again in other works (see for instance \cite{bardi2}). Of course part of human behaviours is not strictly deterministic and depends on cultural factors, which brings about some degree of incertitude in the possibility to predict historical evolutions.

Here I would like to discuss once more the dynamics of \textit{growth }from
a strictly physical point of view. The starting point is the remark that
whatever one means by "economic growth" it has unavoidably a material
basis: in a way or another the \textit{growth} implies an increase of the
amount of matter manipulated and transformed, then an increase of the amount
of energy used to transform and move matter around. In the rest of the paper
the units, be it explicitly or implicitly, will be physical units, such as
kilograms, joules and so on. These units are unaffected by price movements.
No trend discussed in the following will be expressed in terms of money,
since the latter is \textit{de facto }a human convention regulating
the reciprocal right of access of humans to goods, services and resources.

Once we decide to stay with matter and energy we must take note that they
are governed and constrained by laws which are not decided by parliaments or
dictators, are not affected by the ups and downs of stock exchanges nor are
they sensible to schools of thought or journalistic comments. Laws
constraining matter and energy are \textit{discovered,} not decided, by
science. For what matters here, the main constraint was formulated by
Lavoisier \cite{lavoisier}, inspired by what he saw in chemical reactions, but somehow
echoing ancient formulations going back to Lucretius ("\textit{...nullam rem e nihilo gigni...}": from nothing comes nothing) \cite{lucretius} and earlier to Empedocles (5th century BC). Wording this constraint as "\textit{Nothing can be created or destroyed; everything is transformed}" we directly include both matter and energy, so
complying with modern relativity which states the equivalence of the two. By
the way, the matter-energy conservation law cannot be overthrown by any
technological progress, because technology applies science, does not
trespass it.

One more precision is in order, concerning the "container" of our
socio-economic system, i.e. our planet. The earth is a closed, non-isolated
system. "Closed" means that the exchange of matter with the outside is
negligible with respect to the total mass. "Non-isolated" means that the earth
exchanges energy with the outside in the form of radiation: the input comes
from the sun (not considering a marginal contribution from the part of cosmic rays that do not
originate from our star); the output is again in the form of radiation
emitted outwardly by the ground and the atmosphere. The approximate balance
between the two fluxes is governed by the laws of thermodynamics treating
the earth as a "grey body".

An additional constraint concerns \textit{trasformations} which are so
important both for matter and for energy. Transformation processes are
governed by the \textit{second principle of thermodynamics} which was
initially formulated in the middle of the $19^{th}$ century referred
to thermal engines, but can be generalized to all processes in complex
systems, relying on Boltzmann's statistical formulation of thermodynamics.
For what matters here, the principle may be colloquially explained as
follows. Whenever you start a physical process aiming at converting energy
into something you deem useful to you (let us call it "work") you never can
transform the initial amount of energy into work completely: there will
always be (even in ideally perfect conditions) some "waste" you will
disperse in the environment. Most often the "waste" will be residual
non-retrievable heat; more generally it will be "disorder" in a form or
another (technically: \textit{entropy}). In a closed and isolated context
the "waste" (which includes ordinary garbage) will accumulate; if you wish
to keep your living space in order, you need to get rid of the "waste"
somehow throwing it out of the window. The way nature expels "disorder"
pursuing new dynamical equilibrium states is by raising the thermodynamical
temperature.\footnote{%
This mechanism should not be confused with the greenhouse effect. The former
produces an increase of the global temperature as seen from outside; the
latter leaves the temperature seen from faraway unchanged but modifies the
temperature profile from the low layers to the high atmosphere.}

Usually the above is considered as having little to do with economics. In what
follows I shall show it has a lot.

\section{Growth in a finite environment}
\label{sec:1}
Let us consider the earth as a container filled up of something I shall call
"primary resources", which include matter in any form and energy as well;
the total amount of "resources" be $\Sigma$. Then let us start some process
converting "primary resources" into "goods", which means anything deemed
useful or of interest to humans (including services, which always have a
material basis). I stress the fact that the mentioned process is not a
peculiar one, but rather the set of all single processes activated everywhere for
specific production chains. The picture is complete when we add a continuous
push toward steadily increasing the quantity of "goods", $G$; leave for the
moment aside any negative feedback or side effect.

The simplest growth dynamics, under these conditions, is the same as that
described by Verhulst for the world population with a finite food
availability. In an elementary time $dt$ the increase $dG$ is proportional
to the existing stock of "goods" (every single existing asset concurs to the
global growth). To say better: the pure proportionality is corrected by a
factor feeling the proximity to the "roof" $\Sigma$ and tending to $0$ while
approaching $\Sigma$. In symbols it is:

\begin{equation}
dG=\alpha \left( 1-\frac{G}{\Sigma}\right) Gdt
\end{equation}

Of course if the "primary resources" are infinite ($\Sigma\rightarrow \infty $)
the relation is a sheer proportionality.

The constant $\alpha $ is the relative initial growth rate. It is convenient
to normalize the quantity of "goods" to the total available stock of
"primary resources" introducing the variable:

\begin{equation}
g=\frac{G}{\Sigma}
\end{equation}

The basic relation then becomes:

\begin{equation}
dg=\alpha \left( 1-g\right) gdt  \label{log}
\end{equation}

As it is well known, integrating (\ref{log}) one obtains a logistic
curve:

\begin{equation}
g=\frac{1}{1+qe^{-\alpha t}}
\label{logistic}
\end{equation}

The constant $q$ is related to the initial value of $g$: $g_{0}=g_{\left(
t=0\right) }$. It is

\begin{equation}
q=\frac{1}{g_{0}}-1
\end{equation}

then

\begin{equation}
g=\frac{g_{0}}{g_{0}+\left( 1-g_{0}\right) e^{-\alpha t}}  \label{wealth}
\end{equation}

When $g_{0}\ll 1$ and we are close to the origin ($t\ll 1/\alpha $) the
trend is similar to an exponential: $g\simeq g_{0}e^{\alpha t}$.

\begin{figure}[ht]
	\begin{center}
		\includegraphics[width=10cm]{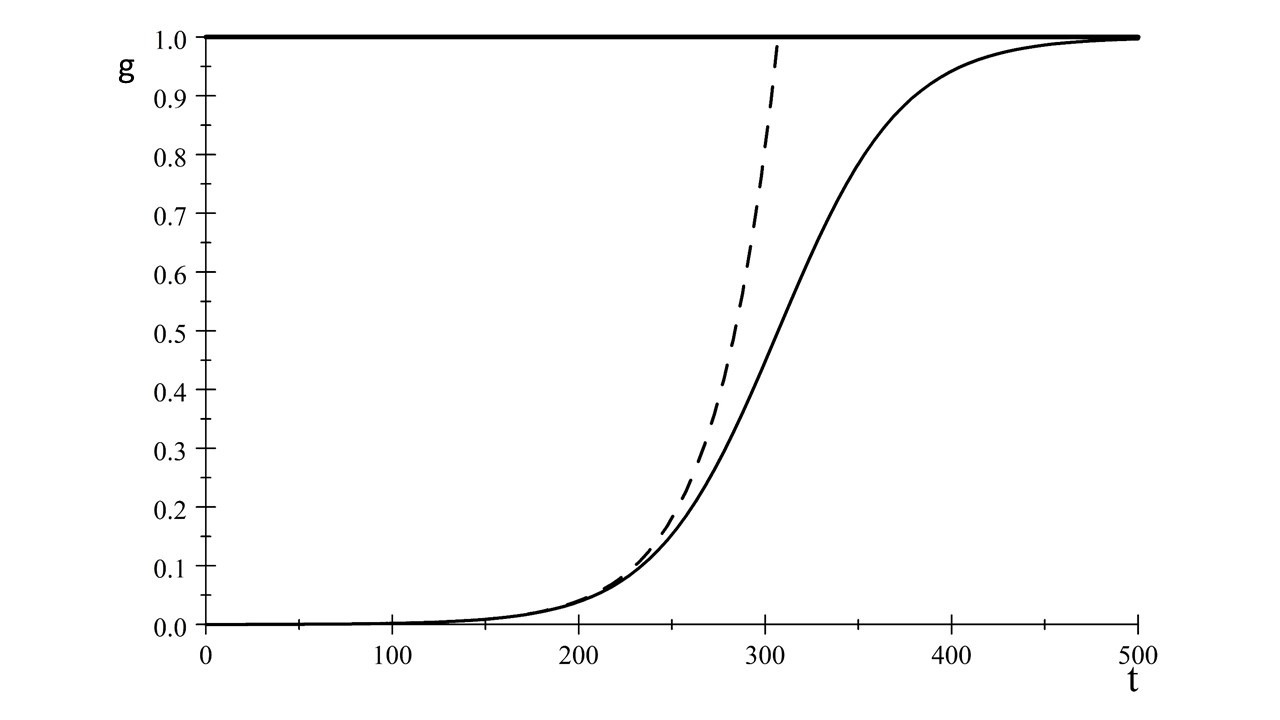}\\
		\caption{Logistic evolution of the
amount of "goods" produced in a finite resources scenario. The dashed line
represents the exponential trend approached in the early phase of the
process. The initial growth rate has been $\alpha =0.03$ per year;
$g_{0}=10^{-4}$.}
		\label{fig:1}
	\end{center}
\end{figure}

Fig. \ref{fig:1} exemplifies the growth evolution I have described.

Just to fix numbers for drawing the graph and without attaching too much
relevance to the choice, I have assumed $g_{0}=10^{-4}$ which is the same
order of magnitude as the ratio between the total present energy consumption
of the human kind and the incoming flux of radiative energy from the sun.
The initial growth rate has been chosen to be $3\%$ per year. The highest
is $\alpha $, the sooner the curve reaches its inflection point (at time $%
t=t_{i}$); it is

\begin{equation}
t_{i}=\frac{1}{\alpha }\ln \frac{1-g_{0}}{g_{0}}
\end{equation}

In the example of Fig. \ref{fig:1} it is $t_{i}=307$ years. For bigger and bigger values of $g_0$ the inflection point moves to the left and when $g_0\geq 1/2$ no inflection point is found between the origin and infinity.

All the above is based on the declared assumption that growth is indeed material, however, in the current debate, a widely used locution mentions a "green growth" based on "decoupling" where GDP (or welfare?) continues to grow indefinitely but, at least in the most radical version of the idea, the implied matter and energy stay fixed. The possibility to have an immaterial growth (under market conditions) resting on globally constant amounts of manipulated matter and transformed energy has no rational content and motivation: it is pure ideology. Furthermore even on empirical bases, the alleged decoupling strategy seems to be contradicted by real economic dynamics \cite{hickel}.

\section{"Costs"}
\label{sec:2}
The dynamics described in the previous section illustrates a principle
situation evidencing a basic mechanism, but is, strictly speaking,
unrealistic, because, as declared, it does not take into account any
back-reaction or side effect. Any physical growth mechanism requires that
part of the primary resources, as well as part of the goods globally
produced, be destined to insure and preserve the efficiency of the
conversion process. You may include in this need maintenance, safety and the
like. I shall call this quota of resources/goods "costs", $C$, reminding
once more that these costs are \textit{not }measured in terms of money and prices but
using physical units.

The importance of "costs" and their specific nature varies according to
different peculiar processes, but here we are interested in considering only the global
conversion system from resources to goods and our attention will focus on general
features governing all processes.

To produce anything we need some amount of raw material and some energy flux that feeds the process. When we wish to increase the production, the demand of raw material increases proportionally, but not so for energy. Energy $E$ is conveyed by some carrier $\Phi$ and in order to increase its flux you need to perform some \textit{work} which is proportional to the change you desire and to the size of the variable you plan to increase. The elementary relation is

\begin{equation}
dE=\beta \Phi d\Phi
\end{equation}
where $\beta $ is an appropriate constant.

Integrating, we obtain

\begin{equation}
E=\frac{\beta }{2}\Phi^{2}
\end{equation}

which means that the required energy flux  grows faster than the flux of the energy carrier.

This rule is general and the examples in the physical world are numerous.
The simplest may be found in mechanics, where kinetic energy $K$ depends
quadratically on the speed $v$ and $\beta $ is the mass $m$ of the object
which moves:

\begin{equation}
K=\frac{m}{2}v^{2}
\end{equation}

Doubling the speed of a mass quadruples its kinetic energy: the cost of a
higher speed, in terms of energy, grows faster than the speed does.

Another simple example is electric current $I$. When a current flows in a
wire part of its energy is converted into waste heat $Q$ in the wire, so
that you have to lose part of the initial energy. The heat is proportional
to the square of the current (Joule's law):

\begin{equation}
Q=\frac{R}{2}I^{2}
\end{equation}

If the current doubles, the energy you have to pour in to compensate for the
waste heat quadruples. Now $R$ is Ohm's resistance of the wire.

Most often people concentrate on $\beta $ trying and reduce it as much as
possible, but disregarding the square law which is the real problem whenever
the energy demand pretends to grow.

Summarizing the above remarks and considering jointly raw materials and energy as "costs" measured in units of the total amount of resources $\Sigma$, the  dynamics of costs as a function of wealth production is expressed by the formula:

 \begin{equation}
 C=\mu g + \beta \frac{g^2}{2}    \label{cost}
 \end{equation}

So far we have seen that Eq. (\ref{cost}) is a general rule for single
physical processes, however there is more when we have to deal with complex
systems. The subject of complexity is indeed a very complicated and subtle one, whenever all human aspects are included (see for instance \cite{tain}); however, as stated more than once, I stay here again on the side of physical reality and physical variables.

Even so, global economy is undoubtedly a very complex system that we can schematize
by a great number of \textit{knots}, i.e. places (factories, workshops,
agencies ...) where primary resources are converted into "goods", and by a
big number of links among the knots along which matter and energy (raw
materials, ware, people...) flow.

For the processes in the knots and the "current" along the links the general
cost law (\ref{cost}) holds, but now another aspect related to complexity
enters the scene \cite{tartaglia}.

A simple way to measure the complexity of a network is to count the
relations or links among the knots. Including all possible connections $r$
we see once more that their number depends quadratically on the number of
knots, $n$:

\begin{equation}
r=\frac{1}{2}n\left( n-1\right)  \label{link}
\end{equation}

The number of actual links does not necessarily coincide with all possible
links (\ref{link}), however in a system which is pushed to grow the trend is
towards saturation of the number of links, then saturation of the flow
across each link and of the production capacity of each knot, finally toward
increasing the number of knots. Growth implies also a growth of complexity,
so, summing up and combining (\ref{cost}), holding for each element, with (%
\ref{link}), holding for the whole network, we infer that the cost to keep
the system working grows \textit{more than quadratically} with respect to
the output of the system:

\begin{equation}
C\geq \mu g + \frac{\beta }{2}g^{2}
\end{equation}

Optimistically staying with the lower limit and recalling Eq. (\ref{wealth})
we explicitly write the time evolution of the global "costs" of a growing system in
a finite environment:

\begin{equation}
C=\mu \frac{g_0}{g_0+(1-g_0)e^{-\alpha t}}  +\frac{\beta }{2}\frac{g_{0}^{2}}{\left[ g_{0}+\left( 1-g_{0}\right)
e^{-\alpha t}\right] ^{2}}  \label{costtime}
\end{equation}

Constants $\mu$ and $\beta $ can be expressed in terms of the initial conditions.
Suppose that at time $0$ the initial cost $C_{0}$ be a given fraction $%
\varepsilon <1$ of the initially available "goods":

\begin{equation}
C_{0}=\varepsilon g_{0}  \label{initial}
\end{equation}

Using (\ref{costtime}) and (\ref{initial}) we get:

\begin{equation}
\varepsilon = \mu + \frac{\beta}{2}g_0
\end{equation}%
For simplicity let me assume that the weight of the two components of "costs" be initially the same: $\mu=\frac{\beta}{2}g_0$; we end up with

\begin{equation}
\beta=\frac{\varepsilon}{g_0} \qquad
\mu=\frac{\varepsilon}{2}
\label{betamu}
\end{equation}

then
\begin{equation}
C=\frac{\varepsilon}{2}\frac{g_0}{g_0+(1-g_0)e^{-\alpha t}} \left(1+\frac{1}{g_0+(1-g_0)e^{-\alpha t}}\right)
 \label{costo}
\end{equation}

\begin{figure}[ht]
	\begin{center}
		\includegraphics[width=10cm]{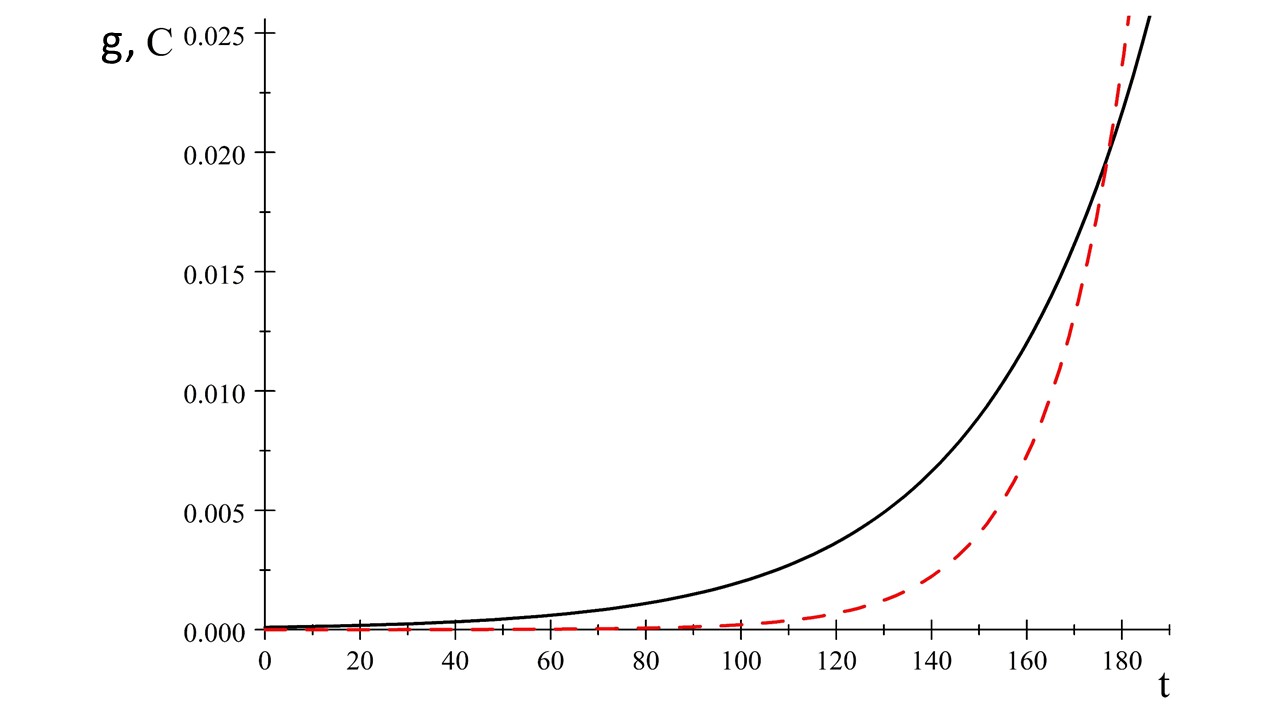}\\
		\caption{The solid line reproduces
the "goods" $g$ and is the same as in Fig. \ref{fig:1}. The dashed
line represents the "costs" $C$ with an initial value equal to $1\%$ of the
"goods".}
		\label{fig:2}
	\end{center}
\end{figure}

What happens with global growth in a closed and of course finite system while time goes on is shown in Fig. \ref{fig:2}%
, drawn using the same numerical values as for Fig. \ref{fig:1} and
assuming that the initial "cost" be $1\%$ of the initial stock of "goods".
The intersection between the two curves happens at time:

\begin{equation}
t_{\ast }=\frac{1}{\alpha }\ln \frac{(2-\varepsilon)(1-g_{0})}{\varepsilon -(2-\varepsilon) (1-g_{0})}
\label{intersect}
\end{equation}

With the data in the example we see that the conflict between production and
costs is reached well before the inflection point of the logistic: here
around year $177$. In general, the intersection exists if $\varepsilon > 2g_0/(1+g_0)$ and costs completely absorb the gross production before the inflection point  whenever $\varepsilon >4g_{0}(1-g_0)/(1+g_0(1-2g_0))$. In practice, in
a growing system the most stringent constraint comes from the costs dynamics
before than from the residual availability of resources.

\section{Benefits}
\label{sec:3}
The discussion in the previous section has started from the fact that part
of the "goods" globally produced must be destined to keep the production
process going. This can also be interpreted saying that the actual
"advantage" or "profit" or "gain" $A$ of the process is the difference
between the gross production $g$ and the necessary cost $C$.

Recalling previous formulae we have:

\begin{equation}
A=g-C=\frac{g_{0}}{g_{0}+\left( 1-g_{0}\right) e^{-\alpha t}}\left( 1-\frac{\varepsilon}{2}-\frac{1}{2}\frac{%
\varepsilon }{g_{0}+\left( 1-g_{0}\right) e^{-\alpha t}}\right)  \label{gain}
\end{equation}

\begin{figure}[ht]
	\begin{center}
		\includegraphics[width=10cm]{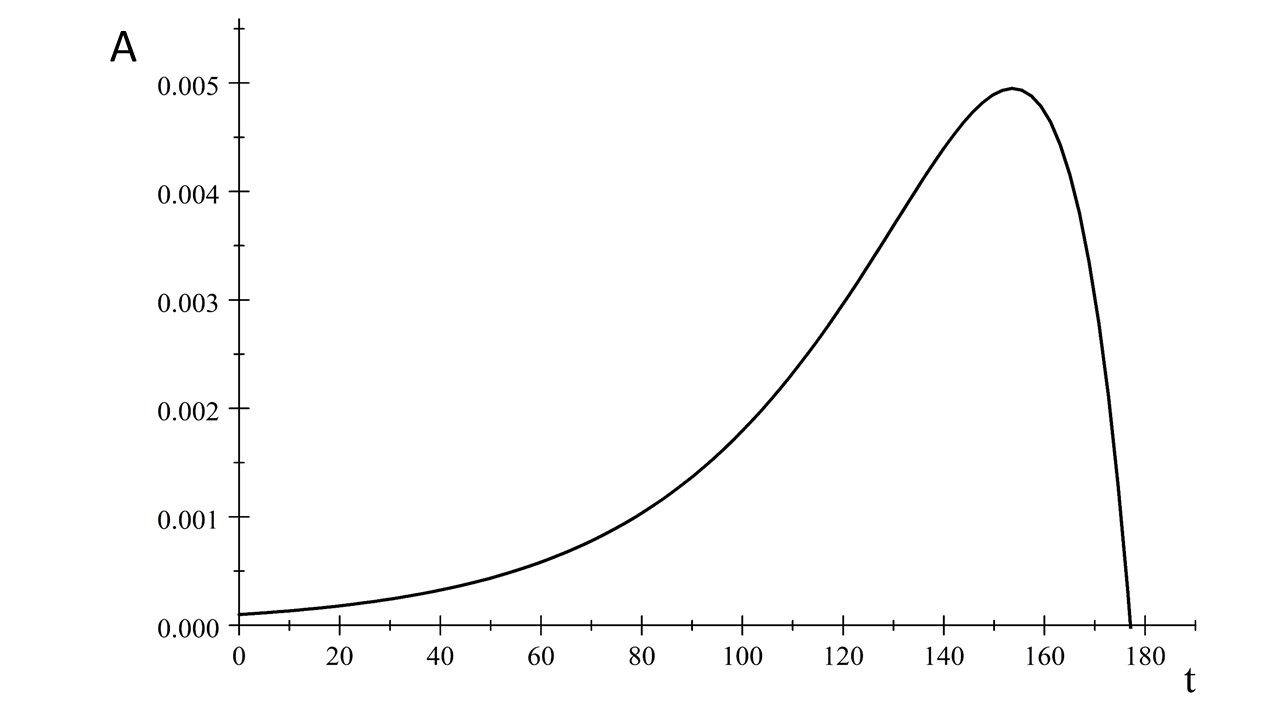}\\
		\caption{Plot of the evolution in
time of "gains" provided by a physical growing system.}
		\label{fig:3}
	\end{center}
\end{figure}

The curve is shown in Fig. \ref{fig:3} and uses the same numerical values
as in the previous graphs. $A=0$ is reached at $t=t_{\ast }$. The maximum is
attained at time:

\begin{equation}
t_{M}=\frac{1}{\alpha }\ln \left(1-\frac{2-3\varepsilon}{(2-\varepsilon)g_0-2\varepsilon}\right)
\label{climax}
\end{equation}

A curve like the one in Fig. \ref{fig:3} is empirically known since a
long time, based on observation of social or personal dynamics or on studies
on human civilizations \cite{diamond}. So much so that it has been nicknamed "Seneca's
cliff" \cite{bardi} from a sentence written by the Roman philosopher Lucius
Annaeus Seneca to his pupil Lucilius: "\textit{incrementa lente exeunt,
festinatur in damnum" }(increases are of sluggish growth, but the way to
ruin is rapid) \cite{seneca}. Examples of rise and fall of historical
civilizations, where the decline is much shorter than the rise, may be found
for instance in Ref. \cite{tainter} where it is interpreted in terms of "diminishing returns".

It is however true that rise and fall not always follow the trend of Fig. \ref{fig:3}. There are examples where the decline happens slowly and the ascent is fast. Most often the collapse does not reach zero: the yield of the system attains a very low level, then stagnates there for some time before witnessing a recovery and starting a new cycle. These differences become manifested when considering more or less local subsystems or peculiar cycles; the less the system is isolated, the more it deviates from the simple trend shown in Fig. \ref{fig:3}. The origin of the deviations is in a plurality of interactions with the rest of the global context and in being the nature of the production process peculiarly linked to some specific kind of resources; the issue has been discussed for instance in Ref.s \cite{taint,bardi2}. Usually people assume that after a collapse a new cycle will follow, based on different assets and techniques; the implicit assumption in this conviction is that there is an infinity of different opportunities. Unfortunately such infinity does not exist: "there is no planet B".

Here I have not considered special cases and specific subsets of the global economy, but the global system as such, and I have highlighted a basic mechanism \textit{necessarily} driving growing closed physical
systems to collapse.

\section{Inequalities}
\label{sec:4}
A recurring worry often recalled and discussed is about income inequalities.
World statistics tell us that inequalities have been increasing everywhere
in the last forty years or so (see for instance Ref. \cite{helpman} and 2018 \textit{World Inequality Report} \cite{ineq}). This trend is present in countries of different continents, with different kinds of government or
regimes, and different governance of the economy, such as USA, on one side,
and China, on the other.

The evolution in different countries is often irregular and noisy depending on
local economical dynamics and expedients and episodical policies aimed at
redistributing income, but the underlying trend looks similar for all countries.

Once more the attention of scholars seems to be concentrated on the empirical description of the dynamics of inequalities in single countries or even globally, looking for expedients or "tricks" to counter the trend they see at work \cite{piketty,atkinson,milanovic}. The question I would like to address here is instead: is there any common mechanism at the base of the generation and increase of differences?

We have already discussed the dynamics of growth which is a central
requirement of globalized economy. The other essential ingredient is \textit{%
competition}, seen as being the main engine of "progress". Let me then try
and analyze, always from a physical viewpoint, the dynamics of competition
in economies striving for growth.

The situation is extremely complicated, but let me reduce the problem to the
essence and consider just two contenders, labelled $1$ and $2$. Two forms of
competition will be discussed. The first one is in a sense a passive
competition: each competitor works to transform primary resources, whose
stock is unique, into "goods"; each competitor wants to grow; each one's
product is of its exclusive pertinence, but the contenders do not directly
interact with one another.

Adapting the approach in (\ref{log}) to the new situation, I write

\begin{equation}
\left\{
\begin{array}{c}
dg_{1}=\alpha _{1}g_{1}\left( 1-g\right) dt \\
dg_{2}=\alpha _{2}g_{2}\left( 1-g\right) dt%
\end{array}%
\right.  \label{system}
\end{equation}

where at any moment it is $g_{1}+g_{2}=g$.
The growth of each player is proportional to its cumulated stock, but is slowed while approaching to the "roof" of the available resources, which is one and the same for both.

In the most general case (any value for $\alpha_1$ and $\alpha_2$) Eq.s (\ref{system}) need to be integrated numerically. However to have an idea of global trends it is enough to consider the special case where $\alpha_1 = \alpha_2 =\alpha$. In such situation, $g_i$ (index $i$ is either $1$ or $2$) and of course $g$ evolve according to a logistic trend. Besides Eq. (\ref{wealth}) we have:

\begin{equation}
g_{i}=\frac{g_{i0}}{  g_{0}+\left( 1-g_{0}\right) e^{-\alpha t} }
\end{equation}

What happens is that the initial difference between
the competitors grows in its absolute value following a logistic curve, but
stays fixed as a fraction of the total amount of "goods": relative
differences are frozen.

If the basic growth rates of the competitors are
different, under the condition of a fixed total amount of primary resources,
the player who has the higher value of $\alpha _{i}$ prevails and continuous to grow following
a trend similar to a logistic curve; the difference between the two also grows approximately logistically. No spontaneous mechanism leads to a long term reduction of the initial difference.

Let us introduce costs and suppose that their initial value be the same fraction $%
\varepsilon $ of the initial stock of "goods" for both contenders. Using condition (\ref{betamu}) it is

\begin{equation}
C_{i}=\frac{\varepsilon}{2} g_i \left( 1+ \frac{g_{i}}{g_{i0}}\right)
\end{equation}%
and the typical "gain" is

\begin{equation}
A_{i}=g_{i}\left[1-\frac{\varepsilon}{2}\left( 1+ \frac{g_{i}}{g_{i0}}\right)\right]
\end{equation}

Each competitor follows its own Seneca's curve; the one who has a higher
initial growth rate meets its collapse earlier than the other.

\subsection{Active competition}
\label{sec:5}
Let me discuss now a more realistic form of active competition: when two contenders compete - i.e. if they do not collude - one tends to get some of the share of the other. The basic equations describing this dynamics can be:

\begin{equation}
\left\{
\begin{array}{c}
dg_{1}=\alpha_1 g_{1}\left( 1-g\right) dt+p \gamma g_{2}dt-(1-p)\gamma g_{2}dt \\
dg_{2}=\alpha_2 g_{2}\left( 1-g\right) dt+ (1-p)\gamma g_{2} dt -p \gamma g_{2}dt
\end{array}%
\right.  \label{winlos}
\end{equation}

The first term on the right accounts for what comes directly from primary resources; the second is what is gained with competition and the third is what is lost. Now $\gamma$ is an empirical (positive) parameter fixing the fraction of the weakest contender's stock of "goods" that can be lost (and gained by the other) during $dt$. The reason to refer to the "goods" of player $2$ amounts to assume that player $1$ is ahead at the beginning and that the stakes are not more than what player $2$ has.
The probability to win for the first contender is $p$ and $1-p$ is the probability to lose. The confrontation is treated as being a continuous process, but we may also think at it as to a series of single episodes in which each party may win or lose. Over a given time $dt$ the cumulated balance of gains and losses is expressed by the statistical effect given by the probability at work, like in Eq.s (\ref{winlos}).

The amount of "goods" which is transferred in one single event of confrontation is assumed to be the same for both players (as in a symmetric bet).

From now on, the path becomes slippery since we are forced to leave the domain of physics and strict rationality and let in elements related to human behaviours. Being aware of this uncertainty I assume the probability to win to be proportional to each player's wealth; in practice it would be

\begin{equation}
p=\frac{g_1}{g}
\label{prob}
\end{equation}

The choice of $g_1$ in the formula is not a forcing since, under the symmetric bet hypothesis, the probability to lose for a contender coincides with the probability to win for the other, and, when it is $p<1/2$, player $1$ is likely to succumb in the confrontation to the advantage of player $2$. The only warning for the calculation is that, in such a situation, you also have to substitute $g_1$ instead of $g_2$ in the last two terms of Eqs. (\ref{winlos}).

In the special case in which $\alpha_1=\alpha_2=\alpha$ the system (\ref{winlos}) has an analytic solution.

Indeed, under the above assumption, we may introduce (\ref{prob}) into (\ref{winlos}), then sum the equations, and put the result in place of the first. Recalling that $g_{1}+g_{2}=g$ the result is:

\begin{equation}
\left\{
\begin{array}{l}
dg = \alpha g\left( 1-g\right) dt \\
dg_2 = \alpha g_2 \left( 1-g\right) dt- \left(1-2\frac{g_2}{g}\right)\gamma g_2 dt
\end{array}%
\right.
\end{equation}

The first equation is once more Eq. (\ref{log}) whose solution is (\ref%
{logistic}). Let us introduce this result into the second equation; we are
left with

\begin{equation}
dg_2=\alpha g_2 \left(\frac{q e^{-\alpha t}}{1+q e^{-\alpha t}}\right)dt - \left(1-2g_2(1+q e^{-\alpha t})\right)\gamma g_2 dt
\label{pw}
\end{equation}

The final solution for both $g_1$ and $g_2$ is:
\begin{equation}
\left\{
\begin{array}{l}
g_{1}=\frac{q+He^{\gamma t}}{(1+qe^{-\alpha t})(2 q+He^{\gamma t})} \\
g_{2}=\frac{q}{(1+qe^{-\alpha t})(2q+He^{\gamma t})}
\end{array}%
\right.
\end{equation}

The two integration constants may be expressed in terms of the initial values $g_1=g_{10}$ and $g_2=g_{20}$. In practice the dependence of $g_1$ and $g_2$ on time is similar to a logistic and their sum is exactly logistic. The player which starts ahead stays ahead and the difference between the two also grows logistically.

\begin{figure}[ht]
	\begin{center}
		\includegraphics[width=10cm]{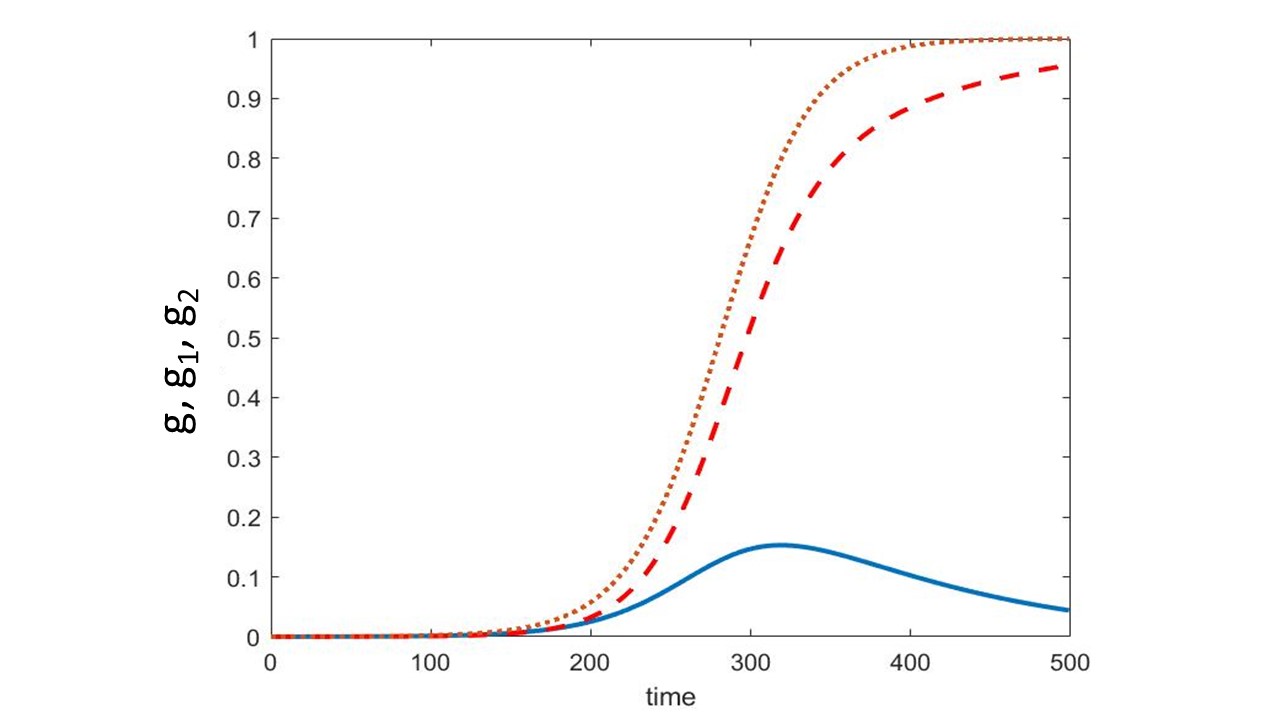}\\
		\caption{Time evolution with different initial growth rates in presence of active competition between two players. The solid line is the
time evolution of $g_{1}$; the dashed line is $g_{2}$. The dotted line is $g$, the sum of
the other two. The numerical values of the parameters may be found in the text.}
		\label{fig:4}
	\end{center}
\end{figure}

It is interesting to consider the case when $\alpha_1< \alpha_2$ and $g_{10}>g_{20}$; in other words the competitor which starts lower has an initial bigger growth rate. The equations must be solved numerically. An example is shown in Fig. \ref{fig:4} where it has been assumed that $g_{10}=0.51\times 10^{-4}$, $g_{20}=0.49 \times 10^{-4}$, $\alpha_1=0.03$ and $\alpha_2$ is $25\%$ higher than $\alpha_1$, $\gamma=0.01$.

What happens is that player $2$ progressively reduces its disadvantage, then surpasses its competitor. From then on its advantage continues to grow steadily, asymptotically tending to a logistical trend.

Different sets of the values of the parameters lead to different scenarios, but in general whenever the "bet", $\gamma g_{lower}$, is less than the amount of growth per unit time, the one who starts expanding at the highest rate takes over and, from then on, continues to increase its advantage. Of course if the highest initial growth rate is with player $1$ the difference with player $2$ inexorably grows from the very beginning.

Let me now include costs as per Eq. (\ref{cost}) and use the simplifying assumption leading to (\ref{betamu}). I also assume that the initial "costs" are the same fraction $\varepsilon $ of $g_{i0}$ for both players. The explicit equation for the "gain" of either competitor is:

\begin{equation}
A_{i}=g_{i}-\frac{\varepsilon}{2}g_{i}\left(1+\frac{g_i}{g_{i0}}\right) \\
\label{gaingraph}
\end{equation}

\begin{figure}[ht]
	\begin{center}
		\includegraphics[width=10cm]{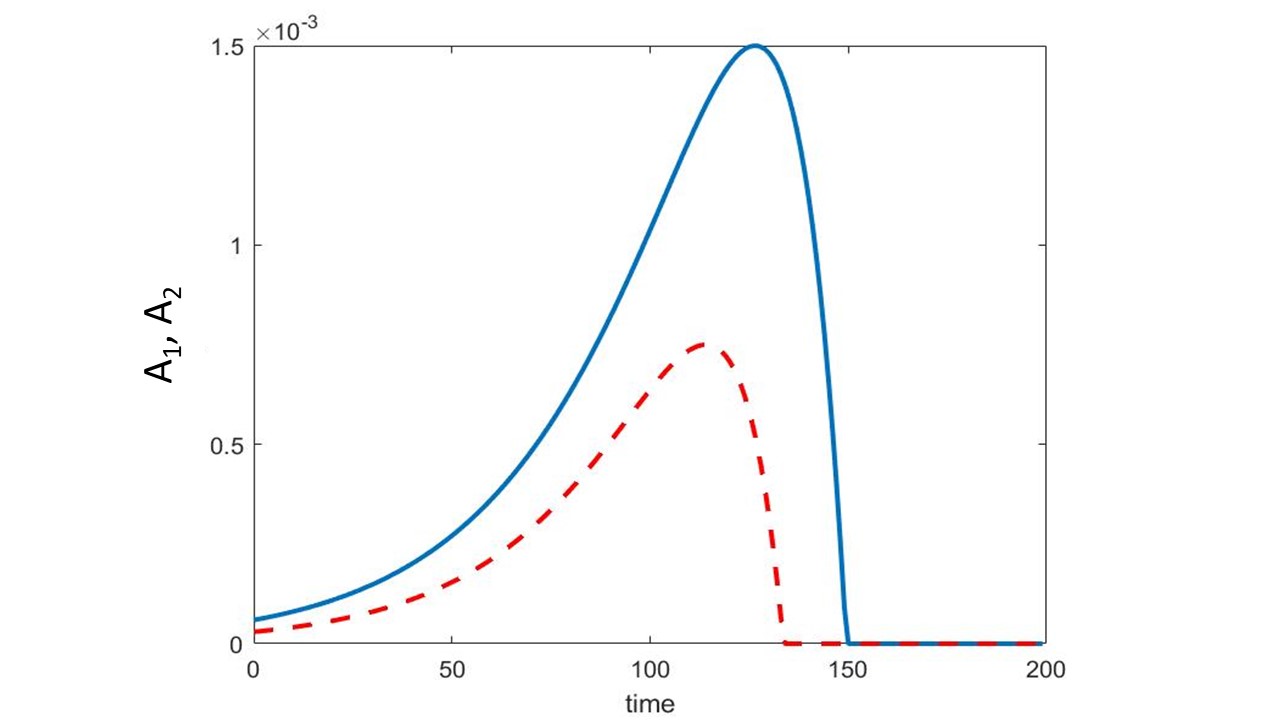}\\
		\caption{The solid line is the
time evolution of $A_{1}$; the dashed line is $A_{2}$. Numerical values of the parameters are the same as in Fig. \ref{fig:4}.}
		\label{fig:5}
	\end{center}
\end{figure}

Reproducing the curves in a graph, we obtain Fig. \ref{fig:5}, where the
numerical values are the same as in Fig. \ref{fig:4}, including $\varepsilon
=0.01$.

As it can be seen, the scenario is dramatically different from the one in Fig. \ref{fig:4}: collapse for both players happens before that the second contender succeeds in passing the first.

Using different sets for the values of the parameters, different results are obtained, but in all cases it can be seen that no spontaneous and automatic mechanism exists leading, with time, to a fair distribution of wealth and of course advantages: in the end inequality always grows.

Having approached the problem from a physical point of view, no negative values have been allowed: nature is not influenced by what does not exist. We however know that it is not always so for humans and their behaviour: the world of finance includes creation and destruction of virtual wealth, even when the existing stock of "goods" remains unchanged. Sometimes a punter may wager something he does not possess and occasionally he may win real riches instead. All these human "miracles" can, now and then, lead to swap the roles between who is upper and who is lower, but the general mechanisms remain the same. The occasional disturbances would appear in the graphs as localised glitches superposed to the smooth trends shown by the graphs and in any case it is clear that no spontaneous mechanism (synthesized as "the market") can automatically produce a fair distribution of wealth.

\section{Conclusion}

The real world with which humans interact is indeed a quite complicated
system: cause/effect relationships are in general non-linear and the
evolution of the whole system is basically chaotic (in the technical meaning
of the word). I have considered extremely simple situations, with the aim of
highlighting the fundamental mechanisms of the machine's operation. Numerical values used in the examples are arbitrary but not entirely irrealistic, so that also the time scale of the figures is plausible.
Those mechanisms are embedded in the real world and all superposed noises and
non-linearities can camouflage them in various ways, but never subvert their
essence and implications. Certain initial conditions invariably produce
certain consequences.

The initial axioms of our globalized economy are two: growth and
competition. The arena in which the global game is played is finite. As we
have seen, the dynamics of production and costs leads the system to collapse
and the details of collapse are irrelevant. Adding competition, we have seen
that the situation does not change as regards the final outcome but in
addition income inequalities grow up to the final collapse phase. This
result has been exposed in terms of equations and graphs, but it is also
perfectly and intuitively exemplified by the Monopoly game: at the beginning
all players are approximately at the same level; in the end the winner has
everything and the others are left with nothing.

If we do not like the ending, we have to change the initial conditions,
intended as the rules of the game. The debate on these issues is normally
encumbered by heaps of political, social, emotional, rhetorical factors,
including some sort of faith in magics and the irrational refusal to look
further beyond the immediate and local context. Unfortunately no irrational
emotionality is able to influence those parts of the rules of the game that
are not under our jurisdiction. This is physics.

\begin{acknowledgements}
I would like to thank Antonino Bonan, Roberto Burlando and Luca Mercalli for reading the manuscript and giving valuable suggestions for its improvement.
\end{acknowledgements}

%
\section*{Conflict of interest}

The author declares that he has no conflict of interest.



\end{document}